\documentclass[Afour,sageh,times]{sagej}
\usepackage{moreverb,url}
\usepackage[colorlinks,bookmarksopen,bookmarksnumbered,citecolor=red,urlcolor=red]{hyperref}

\newcommand\BibTeX{{\rmfamily B\kern-.05em \textsc{i\kern-.025em b}\kern-.08em
T\kern-.1667em\lower.7ex\hbox{E}\kern-.125emX}}

\def\volumeyear{2016}

\usepackage{graphicx}%
\usepackage{multirow}%
\usepackage{amsmath,amssymb,amsfonts}%
\usepackage{amsthm}%
\usepackage{mathrsfs}%
\usepackage[title]{appendix}%
\usepackage{xcolor}%
\usepackage{textcomp}%
\usepackage{manyfoot}%
\usepackage{booktabs}%
\usepackage{algpseudocode}%
\usepackage{listings}%
\newtheorem{theorem}{Theorem}
\newtheorem{definition}{Definition}%
\newtheorem{application}{Application}
\newtheorem{example}{Example}%

\usepackage{hyperref}

\usepackage[utf8]{inputenc}
\usepackage{enumitem}
\usepackage{amsmath}
\usepackage{graphicx}

\usepackage[ruled,vlined,linesnumbered]{algorithm2e} 

\usepackage{graphicx}
\graphicspath{{image/}}

\theoremstyle{definition}

\theoremstyle{definition}
\theoremstyle{definition}
\usepackage{xcolor}

\graphicspath{{image/}}

\newcommand{\adj}{{\it adj}}
\newcommand{\Deg}{{\it deg}} 

\newcommand{\core}{{\it core}}

\newcommand{\pr}{{\it pr}} 
\newcommand{\ch}{{\it ch}} 

\newcommand{\lb}{\it lb}
\newcommand{\lmax}{L_{max}}
\newcommand{\live}{\texttt{Live}}
\newcommand{\dead}{\texttt{Dead}}

\newcommand{\Active}{\texttt{Active}}
\newcommand{\Inactive}{\texttt{Inactive}}

\newcommand{\alg}[1]{\textsc{#1}} 

\newcommand{\pub}{{\it pub}}
\newcommand{\pri}{{\it pri}}
\newcommand{\ver}{{\it ver}}

\newcommand{\myfig}{Figure}

\usepackage{xcolor}
\newcommand{\mytodo}[1]{\textcolor{red}{#1}}

\newcommand{\figscale}{0.95}

\begin{document}

\title{Federated $k$-Core Decomposition: A Secure Distributed Approach}

\runninghead{Guo, Sekerinski, and Chu}
\author{Bin Guo\affilnum{1}, Emil Sekerinski\affilnum{2}, and Lingyang Chu\affilnum{2}}

\affiliation{\affilnum{1}Trent University, Canada\\
\affilnum{2}McMaster University, Canada}

\corrauth{Bin Guo, Trent University} \email{binguo@trentu.ca}

\begin{abstract}
As one of the most well-studied cohesive subgraph models, the $k$-core is widely used to find graph nodes that are ``central'' or ``important'' in many applications, such as biological networks, social networks, ecological networks, and financial networks. 
For Decentralized Online Social Networks (DOSNs), where each vertex is a client as a single computing unit, distributed k-core decomposition algorithms have already been proposed.
However, current distributed approaches fail to adequately protect privacy and security. In today's data-driven world, data privacy and security have attracted more and more attention, e.g., DOSNs are proposed to protect privacy by storing user information locally without using a single centralized server. In this work, we are the first to propose the secure version of the distributed $k$-core decomposition. 
\end{abstract}


\maketitle
\section{Introduction}

Graphs are important data structures that can represent complex relations in many real applications, such as social networks, communication networks, biological networks, hyperlink networks, and model checking networks. 
Given an undirected graph $G$, the \emph{$k$-core decomposition} is to identify the maximal subgraph $G'$ in which each vertex has a degree of at least $k$; the \emph{core number} of each vertex $u$ is defined as the maximum value of $k$ such that $u$ is contained in the $k$-core of $G$~\cite{bz2003,Kong2019}. 
It is well known that core numbers can be computed with linear running time $O(n+m)$ using the \alg{BZ} algorithm~\cite{bz2003}, where $n$ is the number of vertices and $m$ is the number of edges in $G$.
Due to the linear time complexity, the $k$-core decomposition is easily and widely used in many real-world applications.

In~\cite{Kong2019}, Kong et al.~summarize a large number of applications for $k$-core decomposition in biology, social networks, community detection, ecology, information spreading, etc. Especially in~\cite{burleson2020k}, Lesser et al.~investigate the $k$-core robustness in ecological and financial networks.   
In a survey~\cite{Malliaros2020}, Malliaros et al.~summarize the main research work related to $k$-core decomposition from 1968 to 2019.

\subsection{Distributed Graph Models}
In today's data-driven world, data graphs tend to grow continuously and must be distributed rather than stored on a single machine. In particular, data privacy and security have attracted more and more attention~\cite{chen2020survey,terzi2015survey,luo2023survey}. Our work is based on the model for multiple \emph{clients}~\cite{montresor2012distributed}, where each client is a single computational unit like a single machine with shared memory, and different clients communicate by asynchronous network.
By considering privacy and security, we summarize the distributed data graphs into three models:



\begin{itemize}
    \item \textbf{\emph{``One-to-Many''}}:
    The size of data graphs is increasing rapidly in today's data-driven world. For example, the Facebook social network is on the scale of hundreds of billions or even a trillion ($10^{12}$) edges~\cite{ching2015one}. 
    It is impossible to load the entire data graph into the main memory of a single machine. The graph is split into multiple subgraphs and stored in distributed clients.
    Since it is essentially one data graph, it tends to have no privacy or security problems. A centralized server can be used for synchronization and for storing shared data. 

\item \textbf{\emph{``Secure-One-to-Many''}}:
    The data graphs can be inherently distributed. For example, different companies or organizations, e.g., Facebook, Twitter, and the Federal Government, can collaboratively contribute their own data graph as subgraphs for analytics, which must be stored on different independent servers~\cite{li2020review}. Since these subgraphs are always from different organizations, they must protect sensitive information from leaking to other subgraphs when analyzing the entire data graph. 

    \item \textbf{\emph{``Secure-One-to-One''}}:
    In the graph, each vertex can be an independent computation client. 
    For example, mobile phone networks can be modelled as graphs, where users are vertices, and connections between users are edges~\cite{ferrara2014detecting}. Each mobile phone is a computation unit that is inherently distributed. Since nowadays almost everyone has a mobile phone, social networks can be built on such mobile phone networks. Users only know the connection to their neighbours. During computations for graph analysis, users want to protect their privacy and are not willing to leak their information to others, even to their neighbours.  
\end{itemize}

For ``One-to-Many'' and ``Secure-One-to-Many'', vertices of the same client can communicate via shared memory;  
different clients can communicate by passing messages through the network. For ``Secure-One-to-One'', vertices can only communicate by passing messages. 
However, message passing is slower than memory access by an order of magnitude and has high latency~\cite{meng2024survey}. Therefore, we desire a few messages to pass between clients when performing distributed $k$-core decomposition. 


The sequential $k$-core decomposition has been extensively studied~\cite{bz2003,cheng2011efficient,khaouid2015k,wen2016efficient}, and many parallelized algorithms are proposed~\cite{dasari2014park,kabir2017parallel}. 
However, all of these implementations are designed for a single machine, such that the whole data graph has to be stored in the main memory, which cannot handle the above three scenarios. 
To handle distributed data graphs, several distributed $k$-core decomposition algorithms are proposed~\cite{aridhi2016distributed,montresor2012distributed,luo2021distributed,meng2024survey}. 
\iftrue 
The problem is that these traditional distributed $k$-core decomposition algorithms do not adequately consider privacy and security~\cite{parter2019distributed}.
It is urgent to design secure distributed algorithms for the $k$-core decomposition to handle the above ``Secure-One-to-Many'' and ``Secure-One-to-One''.
In this paper, we improve the traditional distributed $k$-core decomposition algorithms by introducing homomorphic encryption~\cite{acar2018survey} to minimize the leakage of vertices' private information, e.g. neighbours and core numbers.  
\fi

For simplicity, in this paper, we mainly focus on the $k$-core decomposition of ``Secure-One-to-One'', assuming that each vertex is a client. It is compatible with ``Secure-One-to-Many'', as the vertices of the same client can also send messages to each other (via memory instead of the network).


\subsection{Data Privacy for $k$-Core Decomposition}
Specifically, for $k$-core decomposition, \emph{privacy} can be defined as follows: a) each client knows its own information; b) each client knows the ID of its directly connected neighbours; c) besides the above two, each client does not know any other information, such as the core numbers of neighbours and the connections of neighbours; d) there does not exist a centralized server for synchronization and storing the information of all clients.

\begin{figure}[tb]
    \centering
    \includegraphics[width=\figscale\linewidth]{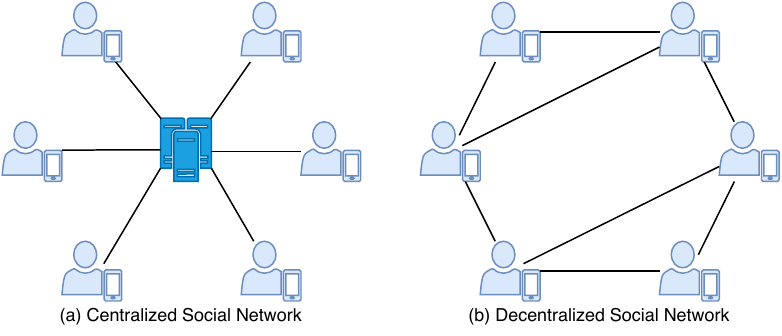}
    \caption{An example of centralized social network vs. decentralized social network.}
    \label{fig:social}
\end{figure}

\begin{application}[Decentralized Social Networks]
Today, the number of Internet users is more than 4 billion, and the number of social media users is about 3 billion. 
A large number of people like to share their daily activities with text, images, and videos on \emph{Online Social Networks} (OSNs), such as Facebook, Twitter, LinkedIn, TikTok, and Instagram. 
In fact, these popular OSNs are centralized as shown in \myfig~\ref{fig:social}(a), which means that the centralized servers store all the user information controlled by a single company. Data on servers can be managed, sold, or stolen without being controlled by the data owner \cite{guidi2020blockchain}. All users may have the risk of their private data being disclosed.

To solve this problem, there are many kinds of \emph{Decentralized Online Social Networks} (DOSNs) \cite{mega2011efficient}, such as Bluesky\footnote{\url{https://bsky.social/}}, mastodon\footnote{\url{https://mastodon.social/}},
Steemit\footnote{\url{https://steemit.com/}}, and Mastodon\footnote{\url{https://joinmastodon.org/}}.
As shown in \myfig~\ref{fig:social}(b), user information is stored locally on distributed networks instead of on a single centralized server. Each user may only know the edges that connect directly to neighbours and does not know any other edges.  
In addition, users may not be willing to share their data to protect their privacy, which can also save the limited bandwidth of clients.
In this case, users can fully protect their private information. 

Our goal is that DOSNs can perform the $k$-core decomposition. 
During the computation, a client needs to compare its core number with its neighbours, without necessarily knowing the accurate core number of its neighbours.
Finally, all clients will obtain their core numbers.
\end{application}

After the $k$-core decomposition, each vertex successfully obtains its core number. 
The other important question is how to release the core numbers. Since core numbers can be private information for clients, it is not safe for all vertices to report their core numbers when there are queries.
We observe that many core number analytics in real graphs~\cite{burleson2020k} only calculate the distribution of core numbers, instead of knowing the specific core number for each vertex. 
That is, each vertex is assigned a label that indicates its special property. We can receive a query, for example, how many vertices have both the label $A$ and the maximum core number $10$, denoted $A{10}$. In this case, essentially, we only need to release the total number of vertices with $A{10}$, and thus we can hide the real core number of each vertex.

\begin{figure}[tb]
    \centering
    \includegraphics[width=\figscale\linewidth]{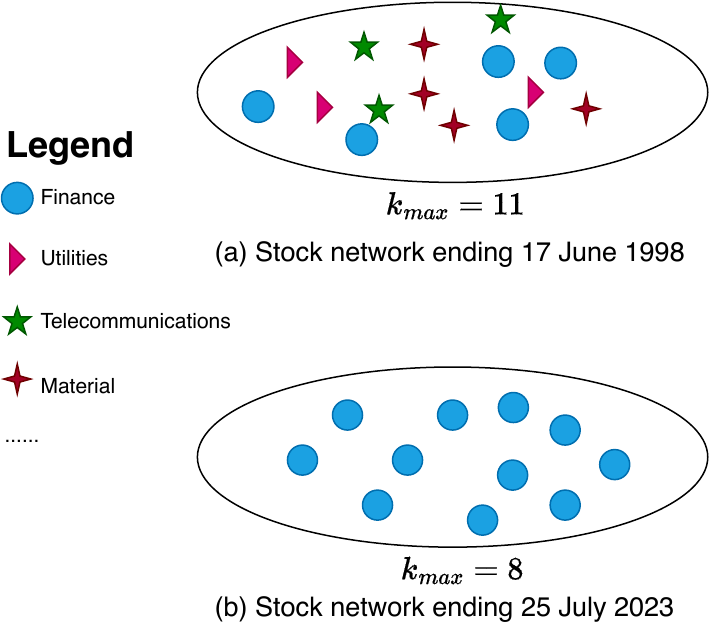}
    \caption{An example of the core number analytics in the stock market networks in different time periods~\cite {burleson2020k}. All the vertices inside the ellipse have the same maximum core numbers, denoted as $k_{max}$. 
    The different shapes and colours of the vertices are their labels, indicating different industries. 
    }
    \label{fig:stock}
\end{figure}

\begin{application} [Stock Market Network]
In \myfig~\ref{fig:stock}, we show an example of core number analytics in stock market networks after $k$-core decomposition~\cite{burleson2020k}. We focus on the vertices with the maximum core numbers. In \myfig~\ref{fig:stock}(a), all vertices with a maximum core number $11$ are located in different industries, e.g., finance, utilities and material. However, in \myfig~\ref{fig:stock}(b),  all vertices with a maximum core number $8$ are dominated by a single industry (Finance), which means that turmoil in the financial sector would have negative effects on the stock market and the economy as a whole. In this case, instead of knowing the exact value of the core number for each vertex, we only analyze the distribution of core numbers with the different labels of the vertices. Therefore, we can control the release of core numbers to protect the privacy of clients.  
\end{application}

However, existing traditional distributed $k$-core decomposition approaches~\cite{aridhi2016distributed,montresor2012distributed,luo2021distributed} inadequately consider the data privacy and security for three problems. 
First, a centralized server connects to all clients to detect the termination $k$-core decomposition, by which the server may know all the IDs of the clients. Second, a client has to store all the core numbers of neighbours, which explicitly exposes its core numbers to neighbours. Third, there are no strategies to safely release the core numbers of vertices after core decomposition. 
In this work, we try to overcome these problems.

\subsection{Contributions}
\begin{figure}[!htb]
\noindent
\centering
\includegraphics[width=\figscale\linewidth]{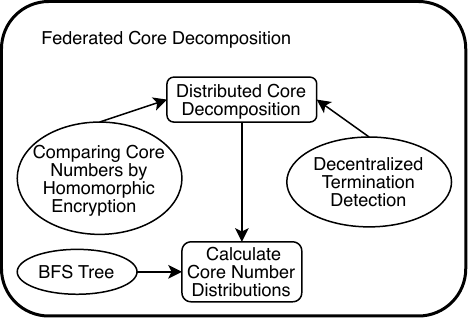}
\caption{The framework of Our Federated $k$-Core Decomposition.}
\label{fig:framework}
\end{figure}

In this paper, we improve the existing distributed $k$-core decomposition algorithm in~\cite{montresor2012distributed} by preserving data privacy and security on ``Secure-One-to-One'', the so-called \emph{federated $k$-core decomposition}.
The general framework is shown in \myfig~\ref{fig:framework}. That is, the distributed core decomposition calculates the core numbers of vertices. The other three approaches are proposed to improve privacy and security for comparing core numbers, detecting termination, and calculating the result of core numbers.
Our main contributions are summarized as follows: 
\begin{itemize}

\item We formally define the privacy of vertices in distributed graphs (in Section \emph{Define the Problem}). 

    \item
We propose Homomorphic Encryption (HE)~\cite{acar2018survey} (in Section \emph{Our Technical Implementation}) to compare a pair of core numbers for two vertices without leaking the value of the core numbers to other vertices. 
That is, two encrypted core numbers can be compared, and we can merely know which one is greater without knowing the exact value.

\item 
We propose decentralized termination detection to identify whether the computation is complete or not (in Section \emph{Our Technical Implementation}). 
The termination condition is that all the vertices are not active together at the same time; that is, {at least one vertex is active in the same time snap}. 
The challenge is that the message passing has a long latency, and it is possible that our detection of ``all vertices are not active'' are not in the same time snap. 
We first build a distributed BFS(Breath-First Searching) tree and then use the heartbeats to detect termination.

\item 
We notice that many applications in~\cite{burleson2020k} (as shown in \myfig~\ref{fig:stock}) only release the distribution of the core numbers for the vertices with the same label. We propose a secure method to count the number of vertices with the same core numbers and labels for safely releasing the results (in Section \emph{Our Technical Implementation}). 
After termination, all the core numbers of vertices have already been calculated. An issue is how to safely release the core numbers. 
We notice that many applications in~\cite{burleson2020k} only release the distribution of the core numbers for the vertices with the same label. For example, we want to know how many vertices a graph has with the label $A$ and the core number $1$. In other words, it is not necessary to know the specific core number for each vertex, which can be a privacy concern, when releasing the core number distribution. To achieve this, by choosing a root vertex, we use the distributed BFS tree to collect the total number of vertices with a specific pair of core number and label. The counter is encrypted by HE. In the end, only the root vertex in the BFS tree has the accumulated counter for the core number distributions. All other vertices only know that one root vertex launches the counting process, but they do not know which pair of core number and label is counted, or even the total or partial counted results. 

\end{itemize}

\iftrue
The rest of this paper is organized as follows. The related work is discussed in Section \emph{Related Work}. 
The preliminary is given in Section \emph{Prilimininaries}. 
Our problem is formally defined in Section \emph{Define the Problem}.
The details of the proposed methods are described in Section \emph{Our Technical Implementation}.
The conclusion and future work are discussed in Section \emph{Conclusion and Future Work}.
\fi

\section{Related Work}\label{relatedwork}
\subsection{The $k$-Core Decomposition} 
In~\cite{bz2003}, Batagelj et al. propose a linear time $O(m+n)$ algorithm for $k$-core decomposition, the so-called BZ algorithm. 
\iftrue 
The general idea is peeling. 
To calculate the $k$ core $G_k$ of $G$, it repeatedly removes those vertices (and their adjacent edges) whose degrees are not larger than $k$. When there are no more vertices to remove, the resulting graphs are the $k$-core of $G$. The peeling process starts from $k=1$ until no vertices and edges can be removed again.

The min-priority queue $Q$ is used to find the vertex of the minimum degree and can be efficiently implemented by bucket sorting, resulting in a linear running time of $O(m)$~\cite{bz2003}, where $m$ is the number of vertices. However, the BZ algorithm requires random access to the entire graph to maintain the queue $Q$. Therefore, all vertices and edges must be stored in main memory, which is essentially a centralized algorithm.  
\fi 
Based on the BZ algorithm, the parallel computation of core decomposition in multicore processors is proposed in ~\cite{dasari2014park,kabir2017parallel}. 

\iftrue 
Based on the BZ algorithm, the parallel computation of core decomposition in multicore processors is first investigated in~\cite{dasari2014park}, where the ParK algorithm was proposed. Based on the main idea of ParK, a more scalable PKC algorithm has been reported in \cite{kabir2017parallel}. Such parallel algorithms are also centralized; the entire graph must be loaded into the shared memory by multicore processors.
\fi 
In~\cite{montresor2012distributed}, Montresor et al. first propose a distributed $k$-core decomposition algorithm that can handle distributed graphs. 
In~\cite{luo2021distributed}, Luo et al. extend distributed $k$-core decomposition algorithms to probabilistic graphs. 
In~\cite{chan2021distributed}, Chan et al. study the approximate distributed $k$-core decomposition. 
In~\cite{liao2022distributed}, Liao et al. explore the distributed $D$-core decomposition specifically in directed graphs; here, $D$-core, also called $(k,l)$-core, is a directed version of $k$-core, the maximal directed subgraph such that every vertex has at least $k$ in-neighbours and $l$ out-neighbours. 
These traditional distributed approaches never consider privacy and security during computation. 

\iftrue 
\subsection{The $k$-Core Maintenance}
The above $k$-core decomposition is for static graphs, which is not changed after the graphs are generated.
In real applications, e.g. social networks, the graphs can be dynamic for continuous changing, since many edges can be frequently inserted or removed.
For each edge insertion or removal, it is time-consuming to recalculate the core numbers after the $k$-core decomposition; instead, we can maintain the core numbers by traversing a minimal size of vertices, the so-called $k$-core maintenance algorithms. 
Many sequential maintenance algorithms~\cite{Saryuce2016,Zhang2017,guo2022simplified} are proposed to maintain core numbers when a large number of edges are inserted or removed. 

Based on these sequential maintenance algorithms, many parallel maintenance algorithms~\cite{hua2019faster,Jin2018,wang2017parallel,guo2023parallel} are proposed for the multicore shared memory architecture, and the one in~\cite{guo2023parallel} is the most efficient in the state of the art. 
Additionally, the work in~\cite{liu2022parallel,liu2024parallel} studies approximately maintaining the core numbers in parallel by guaranteeing the approximation and parallel costs.

In~\cite{mandal2017distributed}, Mandal et al. design the distributed $k$-core decomposition algorithm running on the Spark cluster computing platform. 
In~\cite{aridhi2016distributed}, Aridhi et al. propose distributed $k$-core decomposition and maintenance algorithms in large dynamic graphs, which are partitioned into multiple subgraphs and each of them assigned to a client. In each subgraph, a centralized $k$-core decomposition algorithm is executed to compute the core numbers, which will cause the update of other subgraphs. There is a Master to manage all the clients for communication.
In~\cite{weng2021efficient}, Weng et al. study the distributed $k$-core maintenance algorithms on dynamic graphs. 
\fi  

\iftrue 
\subsection{Weighted Graphs}
All above research focus on unweighted graphs, but graphs are weighted in a lot of realistic applications.
For an edge-weighted graph, the degree of a vertex is the sum of the weights of all its incident edges.
In~\cite{zhou2021core,liu2020Increamental}, the $k$-core decomposition and maintenance algorithms are extended to weighted graphs. 
\fi 

\subsection{Federated Learning}
It is a burgeoning machine learning scheme. The aim is to tackle the problem of the data island while preserving data privacy at the same time. 
The traditional distributed machine learning is to connect multiple clients, e.g. mobile devices, corporations, and institutions, via the network to train a model, which is controlled by a central server and the data privacy is not protected.  
Based on such distributed learning, federated learning is designed to ensure that their data remains decentralized and protected. 
In other words, each client provides its private data to train a model together without leaking its privacy~\cite{mcmahan2017communication, li2020review}.
The behind idea can be applied to distributed $k$-core decomposition algorithms, which is the main contribution in this paper.

\subsection{Other Techniques}
Recently, data privacy and security have been widely studied in different areas, e.g., blockchain~\cite{wang2022bsif}, computer networks~\cite{wang2025efficient,wang2025enhanced}.


\section{Preliminaries}
\label{Preliminary}
Let $G = (V, E)$ be an undirected unweighted graph, where $V(G)$ denotes the set of vertices and $E(G)$ represents the set of edges in $G$. When the context is clear, we will use $V$ and $E$ instead of $V(G)$ and $E(G)$ for simplicity, respectively.
As $G$ is an undirected graph, an edge $(u, v)\in E(G)$ is equivalent to $(v, u)\in E(G)$. 
The set of neighbours of a vertex $u \in V$ is defined by $u.\adj = \{v \in V: (u, v) \in E\}$.
The degree of a vertex $u\in V$ is denoted by $u.\Deg = |u.\adj|$.
We denote the number of vertices and edges of $G$ by $n$ and $m$, respectively, in the context of analyzing time, space, and message complexities.
We denote a set of messages for the communication of vertices by $M = \{m_1, m_2, m_3, ...\}$.  

Furthermore, each vertex $u\in V$ has a label to indicate the local feature, which is defined by $u.\lb$. 
The set $\mathcal{L}$ includes all the possible labels that $u\in V$ may have, denoted $\mathcal{L} =\{u.\lb \mid u\in V\}$.


\begin{definition} [$k$-Core]
Given an undirected graph $G=(V, E)$ and a natural number $k$, a induced subgraph $G_k$ of $G$ is called a $k$-core if it satisfies: (1) for $\forall u \in V(G_k)$, $u.\Deg \geq k$, and (2) $G_k$ is maximal. Moreover, $G_{k+1} \subseteq G_k$, for all $k \geq 0$, and $G_0$ is just $G$. 
\end{definition}

\begin{definition}[Core Number]
\label{def:corenumber}
Given an undirected graph $G=(V,E)$, the core number of a vertex $u\in G(V)$, denoted $u.\core$, is defined as $u.\core = max\{k: u \in V(G_k)\}$. That means $u.core$ is the largest $k$ such that there exists a $k$-core containing $u$.
\end{definition}


\begin{definition} [$k$-Core Decomposition]
Given a graph $G=(V,E)$, the problem of computing the core number for each $u \in V(G)$ is called $k$-core decomposition. 
\end{definition}


\subsection{The Client Connection Model}
In this paper, we deal with distributed algorithms in the \emph{asynchronous network}\cite{awerbuch1985complexity}. 
It is a point-to-point communication network for each \emph{client}, where a client is an independent computational unit. No common memory is shared by the clients, and each client has a distinct ID. Each client can perform local computations, send messages to its neighbours, and receive messages from its neighbours. All messages are passed through a reliable network that has a fixed length to carry a bounded amount of information.

We assume that all clients are reliable and never lose connections, and the message passing between clients has an unpredictable latency at most $D_{max}$ time. There is no shared global clock for synchronization. 

\iftrue 
The work in~\cite{montresor2012distributed} only consider two computational models \cite{montresordistributed}: 
\begin{itemize}
    \item ``One-to-one''. One client is associated with one vertex in the graph. Communication occurs only through direct messages between nodes connected through an edge. 

    \item ``One-to-many''. One client stores multiple vertices together with their local and remote edges. Such a client can be a single machine that can store many vertices efficiently communicating via shared memory. In addition, the client can include multiple machines connected by a high-speed local network, the so-called hierarchical network. 
    This client can first run the local computation and then notify other connected clients by message passing.   
\end{itemize}

The above existing two models do not include privacy and security. In this paper, we focus on secure models. 
For simplicity, we only design our federated $k$-core decomposition algorithm on the ``Secure-One-to-One'', as it can still fit for ``Secure-One-to-Many''. To do this, we suppose that the message passing between vertices in the same clients via shared memory, which has almost no latency. 
\fi 

\subsection{Distributed $k$-Core Decomposition Algorithm}

Our federated core decomposition is based on the distributed core decomposition algorithm in \cite{montresor2012distributed}. 

\begin{theorem}[Locality \cite{montresor2012distributed}]
For all $u\in V$, its core number, $u.\core$, is the largest value $k$ such
that $u$ has at least $k$ neighbours that have core numbers not less than $k$. Formally, we define $u.\core = k$, where
$k \leq   |\{v\in u.\adj: v.\core \geq k\}|$ and $ (k+1) > |\{v\in u.\adj: v.\core \geq (k+1)\}| $.
\label{th:locality}
\end{theorem}

Theorem \ref{th:locality} shows that the vertex $u$ is sufficient to calculate its core number from the neighbours' information. The algorithm is reorganized in Algorithm~\ref{alg:core-decompotion}. 
\begin{enumerate}
    \item For the initialization stage (lines 1 to 4), each $u\in V$ has its \emph{estimate core numbers} initialized as its degree (line 2). Each $u$ will maintain a set of estimated core numbers for all neighbours $u.\adj$ (line 3). Then, each $u$ sends its estimated core numbers $u.\core'$ to all neighbours (line 4), where \texttt{Send}$_u(v, \langle ... \rangle)$ procedure (executed on the vertex $u$) is to send the message $\langle ... \rangle$ to the target vertex $v$ (lines 4 and 11).  
    
    \item Then $u$ will receive the estimated core numbers $v.\core'$ from its neighbours $v\in u.\adj$ (lines 5 to 11). The received $\core'$ will be stored in the array $u.A$ (line 6). If $u.A$ does not include all the estimated core numbers of $u.A$, it will return directly since $u$ has not yet received all the neighbours' messages (line 7); otherwise, $u$ will get the new core number $k$ by executing \texttt{GetCore} (line 8). 
    If $k$ is not equal to $u.\core'$, $u$ will update its core number and send it to all neighbours (lines 9 - 11).
    The \texttt{GetCore} procedure will check the core number with Theorem~\ref{th:locality}; if it is not satisfied, it will decrease the core number $k$ until Theorem~\ref{th:locality} is satisfied (lines 12 - 15). 
    
    \item Each vertex $u\in V$ executes the above distributed algorithm in parallel. The termination condition is that all vertices $u\in V$ satisfy Theorem~\ref{th:locality} and thus stop decreasing the estimated core numbers $u.\core'$ at the same time. Finally, we obtain the final calculated core numbers $u.\core = u.\core'$ for all $u\in V$. 
\end{enumerate}

\iftrue  
\begin{algorithm}[tb]
\caption{Distributed Core Decomposition on Each Vertex $u\in V$ 
}
\label{alg:core-decompotion}
\SetAlgoNoEnd
\DontPrintSemicolon
\SetKwFunction{Initialize}{Initialize$_u$}
\SetKwFunction{Send}{Send$_u$}
\SetKwFunction{Receive}{Receive$_u$}
\SetKwFunction{GetCore}{GetCore}
\SetKw{Goto}{goto}

\SetKwProg{myproc}{procedure}{ \Initialize{}}{}
\myproc{}{

           $u.\core' \gets u.\Deg$\;
    $u.A \gets $ an empty array storing neighbours' core numbers\; 

       \lFor{$v\in u.\adj$}{
            \Send{$v, \langle u, u.\core' \rangle$}
       }
}

\medskip
\SetKwProg{myproc}{procedure}{ \Receive{$ \langle v, \core'\rangle$}}{}
\myproc{}{
    $u.A[v] \gets \core'$\;
    \lIf{$|u.A| < u.\Deg$}{\Return}
    $k \gets $\GetCore{$u.A, u.\core'$}\;
    \If{$u.\core' \neq k$ }{
        $u.\core' \gets k$\;
        \lFor{$v\in u.\adj$ }{
        \Send{$v, \langle u, k\rangle$}
        }
    }

}

\medskip
\SetKwProg{myproc}{procedure}{ \GetCore{$A,  k$}}{}
\myproc{}{


    \While{\rm{ $ |\{i \in A: i \geq k\}| < k$}\label{coredeco:goto}}{
         $k\gets k-1$;
    }

    \Return $k$\;
}

\end{algorithm}
\fi 

\begin{figure}[tbh]
    \centering
    \includegraphics[width=\figscale\linewidth]{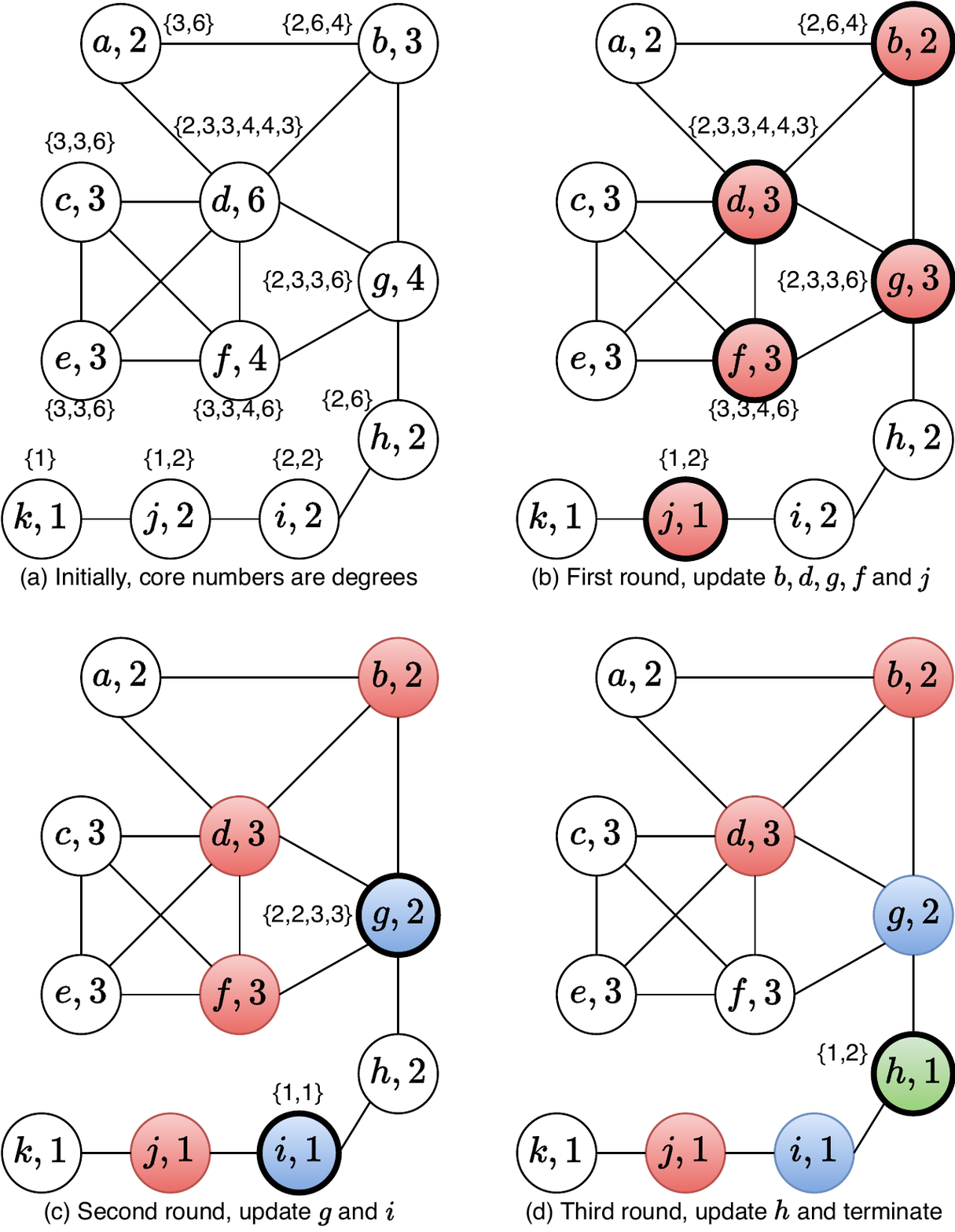}
    \caption{An example graph executes distributed $k$-core decomposition. Inside the circles, the letters are the vertices' IDs and the numbers are the core numbers. The bold circles show that the vertex is active for calculating its core number. The colored circles show that the core numbers have been updated in different rounds. The sets of numbers beside some vertices are their neighbours' core numbers. 
    }
    \label{fig:distributed}
\end{figure}

\begin{example}
In \myfig~\ref{fig:distributed}, we show the $k$-core decomposition in an example graph. \myfig~\ref{fig:distributed}(a) shows that the estimated core number of each vertex is initialized as its degree. For example, the vertex $d.\core$ is set to $6$ since it has $6$ neighbours and its degree is $6$. 
\myfig~\ref{fig:distributed}(b) shows that in the first round, the core numbers for the vertices $b,d,g,f$ and $j$ are active and decrease to $2,3,3,3$ and $1$ (colored red with bold circles), respectively. For example, the vertex $d.\core$ decreases from $6$ to $3$, as $d.\adj$ has a set of core numbers $\{2,3,3,3,4\}$ and $d.\core$ can be at most $3$ since there are at most $3$ neighbours that have core numbers equal to or greater than $3$.
\myfig~\ref{fig:distributed}(c) shows that in the second round, the core numbers for the vertices $g$ and $i$ decrease to $2$ and $1$ (colored purple with bold circles), respectively.
\myfig~\ref{fig:distributed}(d) shows that in the third round, the core number of the vertex $h$ decreases to $1$ (colored green). Finally, the algorithm terminates, since all vertices are inactive and cannot continually decrease their core numbers. As we can see, the core numbers are calculated in three rounds, and the vertex $g$ updates the core number twice.
\end{example}

\iftrue

\paragraph{Message Complexities}
Since most of the running time is spent on the message passing for the distributed $k$-Core decomposition algorithm, we should analyze \emph{message complexities} rather than the time complexities, where the message complexities is to count the number of messages passing between different vertices.   
We analyze the message complexities of Algorithm~\ref{alg:core-decompotion} in the standard \emph{work-depth} model. 
The \emph{work}, denoted as $\mathcal W$, is the total number of operations that the algorithm uses.
The \emph{depth}, denoted as $\mathcal D$, is the longest chain of sequential operations. 

In Algorithm~\ref{alg:core-decompotion}, the work $\mathcal W$ is the total number of messages that the degree reduces to the core number. 
Each vertex must send messages to notify all neighbours whenever its degree decreases by one until its degree is reduced to its core number.
So, we have the work denoted as $\mathcal W = O[\sum_{u\in V}{u.\Deg \cdot (u.\Deg - u.\core)}]$.  

In the worst case, the process can be reduced to sequential running, e.g., a chain graph. In other words, the whole process needs the worst-case $n$ rounds to converge one by one in a chain. We suppose that each vertex is a client and only has one worker, so it cannot send messages to all neighbours in parallel. 
Therefore, the depth $\mathcal D$ is equal to the work $\mathcal W$. 
However, real graphs, for example, social networks and communication networks, are not chain graphs and tend to be \emph{small-world} graphs. They have the property that most vertices are reachable from any other node through a small number of steps, even if the network is large~\cite{newman2003structure}, which exhibits two key properties, high clustering and a short average path length.
Therefore, these graphs have a set of vertices $D$ in the longest chain, which is much smaller than the number of vertices $n$, denoted as $|D| \ll |n|$. For example, on small-world social networks we always have $|D|$ less than $100$ and $n$ can be as large as millions. Different chains can execute in parallel.
Therefore, on average, the depth $\mathcal D= O[\sum_{u\in D}{u.\Deg \cdot (u.\Deg - u.\core)}]$

\fi

\subsection{Termination Detection}
For distributed $k$-core decomposition, we need to detect when the correct core numbers have been reached for all vertices. That is, all vertices stop the calculation of core numbers (including sending and receiving messages) at the same time. In~\cite{montresor2012distributed}, several approaches are discussed: 

\begin{itemize}
    \item 
    \emph{Centralized}: Each client sends heartbeats to the server during the computation. The server is centralized to monitor the whole graph. However, the problem is that the centralized server may leak the privacy of clients. 
    
    \item 
    \emph{Fixed Number of Rounds}: it is a simple way based on the observations that most of the real-world graphs can be completed in a very small number of rounds, i.e. less than ten rounds. Even if the convergence is not reached after very few rounds, the estimation error for the core numbers is extremely low. However, the problem is that the core numbers may not be completely calculated. 

    \item 
    \emph{Decentralized Gossip-based protocols}~\cite{jelasity2005gossip}: each node in the network randomly selects a few neighbours to exchange information with at regular intervals. 
    This randomness ensures that the information spreads quickly and broadly, much like a rumour spreading through word of mouth. Such a decentralized approach does not have a central server and can protect privacy. 
    However, the details of the implementation are not given in~\cite{montresor2012distributed}. 
\end{itemize}

In this paper, we propose a new method for termination detection while protecting the privacy of clients. 

\subsection{Asymmetric Encryption}
Asymmetric encryption, also known as public-key cryptography, is a fundamental technique in the field of cryptography, such as RSA, ECC, and DSA~\cite{zhang2021overview}. This method generates a pair of keys: \emph{public key} and \emph{private key}, which together enable secure communication over untrusted networks:
\begin{itemize}
    \item Public Key $\pub$: it is openly shared and can be distributed widely. It is used to encrypt messages or data intended for a specific recipient.
    \item Private Key $\pri$: it is kept confidential by the owner. It is used to decrypt the data that is encrypted with the corresponding public key. The security of the private key is crucial to the integrity of the cryptographic system.
\end{itemize}

For encryption, the sender acquires the recipient’s public key $pub$; using $pub$, the sender encrypts the message $m_1$ to produce the ciphertext $c$ with the function $E_{pub}$, denoted $c = E_{pub}(m_1)$. 
For decryption, upon receiving the encrypted message, the recipient uses their private key $pri$ to decrypt it with the function $D_{pri}$, denoted $m = D_{pri}(c)$.
Unlike symmetric encryption (the same key for both encryption and decryption), there is no need for a secure exchange of keys. The public key can be distributed openly without compromising security. In this work, such public-key cryptography can be used to safely distribute the key $k$ for symmetric encryption, e.g. our homomorphic encryption~\cite{zhang2021overview} to protect the privacy.

\subsection{Homomorphic Encryption}

\emph{Homomorphic Encryption} (HE)~\cite{acar2018survey} is an encryption method that allows arithmetical computations to be performed directly on encrypted or ciphered text without requiring any decryption. 
In other words, HE allows data processing without disclosing the actual data.
The homomorphic over an operation $\star$ is denoted as $$E_k(m_1) \star E_k(m_2) = E_k(m_1 \star m_2), \forall m_1, m_2\in M$$
where $E_k()$ is the encryption function with the symmetric key $k$ for both encryption and description; both $m_1$ and $m_2$ are possible messages in the set $M$. 

Different HE attempts can neatly be categorized under three types of schemes with respect to the number of allowed operations on the encrypted data as follows:
\begin{itemize}
    \item \emph{Partially Homomorphic Encryption} (PHE) allows only one type of operation with an unlimited number of times (i.e., no bound on the number of usages).
    \item \emph{Somewhat Homomorphic Encryption} (SWHE) allows some types of operations a limited number of times.
    \item \emph{Fully Homomorphic Encryption} (FHE) allows an unlimited number of operations for an unlimited number of times.
\end{itemize}

The integer comparison via HE is denoted as $E_k(m_1) > E_k(m_2) = E_k(m_1 > m_2)$. 
This classical secure integer comparison is one of the first problems introduced in cryptography, the so-called Millionaires' problem~\cite{yao1982protocols}, which discusses two millionaires, Alice and Bob, who are interested in knowing which of them is richer without revealing their actual wealth. 
Many efficient homomorphic comparison methods are proposed~\cite{cheon2020efficient,chakraborty2022efficient}. 
Especially, in~\cite{bourse2020improved}, Bourse et al. propose an efficient modified FHE system for the secure integer comparison. 
In this paper, we can use this HE integer comparison to compare the core numbers of two vertices without leaking the private core numbers of the vertices, when doing the distributed $k$-core decomposition.

The addition of integers through HE is denoted as $E_k(m_1) + E_k(m_2) = E_k(m_1 + m_2)$. In~\cite{van2010fully}, the Dijk et al. propose the FHE scheme. In this paper, we can use this HE integer addition to calculate the total number of vertices with the same label and core number, without leaking the intermediate values to the other unrelated vertices. 

The above comparison and the addition via HE are symmetric versions. They are possible to transform to asymmetric versions, as demonstrated by Rothblum~\cite{rothblum2011homomorphic}. 
That is, the integer comparison via asymmetric HE is denoted as $E_{\pub}(m_1) > E_{\pub}(m_2) = E_{\pub}(m_1 > m_2)$, and we can get the result by $D_{pri}(E_{pub}(m_1 > m_2))$.
In this work, it is used to compare the core numbers while protecting privacy. 
Similarly, the addition of integers via asymmetric HE is denoted as $E_{\pub}(m_1) + E_{\pub}(m_2) = E_{\pub}(m_1+m_2)$, and we can get the result by $D_{pri}(E_{pub}(m_1 + m_2))$.
In this work, it is used to count the number of vertices (with the same core numbers and labels) as a releasing result.

Additionally, almost all the well-known cryptosystems, including HE, are deterministic~\cite{sen2013homomorphic}. This means that for a fixed encryption key, a given plaintext will always be encrypted into the same ciphertext under these systems.
That is, we can compare a pair of integers that are equal or not via asymmetric HE directly by comparing the ciphertext, without revealing their values to each other, denoted $E_{\pub}(m_1) = E_{\pub}(m_2) \equiv m_1 = m_2$. In this work, it is used to compare two vertices if they have the same core numbers and labels, when releasing the results.

\subsection{Distributed Broadcast}
The \emph{broadcast} operation is to send a message from a single vertex to all other vertices in the graph~\cite{erciyes2013distributed}. 
The idea is straightforward: 1) the chosen initial root vertex $r$ sends a message $m_1$ to all its neighbours; 2) any vertices that receive $m_1$ for the first time will forward it to their neighbours that have not yet received $m_1$. A breadth-first search tree (BFS) can be generated, in which each vertex $u$ has one parent vertex defined as $u.\pr$ and multiple child vertices defined as $u.\ch$. That is, $u$ receives the message $m_1$ from $u.\pr$ and sends it to all $u.\ch$ to build the BFS tree. An example BFS tree is shown in \myfig~\ref{fig:bfs-tree}. 


The BFS tree can be used to broadcast the messages starting from the root, the so-called \emph{BFS broadcast}. 
In this paper, such a broadcast operation is used to inform that all vertices in the graph start the core decomposition algorithm. This technique is also used for termination detection and for calculating the core number distribution.


\section{Define the Problem}
\label{problem-define}
In this section, we first define our distributed graphs. Then, we define our privacy and security when executing the distributed core decomposition algorithm. 

\subsection{Graphs Definition}
For simplicity, we specifically define the undirected distributed data graph $G=(V,E)$, where each vertex $u\in V$ is a client, for distributed core decomposition with the following properties:
\begin{itemize} 
\item \emph{``Secure-One-to-One''}:
Each vertex $u$ in the graph $G$ is represented by a client as an independent computational unit. Each vertex $u$ can only communicate with neighbours by passing messages through the network. Each client $u$ has one single worker. That is, each client sequentially executes the program without parallelism. Each vertex $u$ has privacy and security issues during the computation. 

    \item \emph{Static}: 
    The graph $G$ is a \emph{static graph}, that is, $G$ is not changed during the calculation. There are no edges that are inserted or removed, and no vertices can be offline.
    The distributed network may not be reliable. In this work, we assume that some clients can be offline but recover quickly in a short time, like 10 ms.

    \item \emph{Message Passing Latency}:
    For each edge $(u,v)$ in the graph $G$, $u$ and $v$ communicate by sending messages to each other. For each edge, its message passing always has a fixed maximal latency $L_{max}$, and different edges may have different latencies, e.g. 100 or 200 ms. In other words, the graph $G$ can be a weighted graph, where the edges are weighted in the delay time of the message passing. For the Internet, $L_{max}$ is affected by the light speed and transmission equipment, up to hundreds of milliseconds, such as 300 ms\footnote{https://news.yale.edu/2022/05/03/internet-speed-light}; also, $L_{max}$ includes the time that the vertices are offline and recover quickly. We define the maximal latency (see Definition~\ref{def:latency} below). 

    \item \emph{Undirected}: 
    The edges $(u,v)\in E$ and $(v,u)\in E$ are equivalent and have the same properties. The edges $(u, v)$ and $(v, u)$ have the same latency $L_{max}$ for message passing, that is, the direction of the edges does not affect the latency of message passing. 
    
    \item \emph{Reachable}:
    The graph $G$ is \emph{reachable}, that is, each pair of vertices $u, v\in V(G)$ is connected by a path. In this case, the broadcast starting from one vertex can reach all the other vertices in $G$.
    It is true that real-world graphs can be disconnected. One solution is that we can link connected vertices with virtual edges, which ensures that the graph is connected. 

\item \emph{Small Diameter}:
Given a graph $G$, the \emph{diameter} is the longest shortest-path distance between each pair of vertices in a graph $G$, denoted $D(G)$. 
The diameter of the graph $G$ can be small even for a very large graph example, the so-called small world graph~\cite{Hong2013}. The observation is that the vast majority of large graphs in the real world are small-world~\cite{Hong2013}, e.g. social networks.

\item \emph{Message Capacity}: 
Each vertex can send a message of size $O(\log n)$ to neighbours once a time, where $n$ is the number of vertices in the graph $G$. Each vertex has a unique ID of $O(\log n)$ bits. The degree or core numbers of the vertices should be much less than $n$. Practically, $n$ can be $2^{32}$ or $2^{64}$ as an integer. The large message that exceeds the capacity $O(\log n)$ must be split into multiple messages to send and receive. 



\item \emph{False Messages}: We omit all possible adversarial behaviours to simplify the model and focus on the essential part of our approach. That is, all messages are correctly sent and received, and no other attacking fake messages (outside of the system) can be sent and received, e.g. false heartbeat messages do not exist for termination detection in Section 5.2; all clients correctly execute algorithms and pass messages to neighbors, e.g. no vertices can send the false core numbers for comparative in Section 5.1. 


\end{itemize}

\begin{definition}[Maximal Latency]
    Given an edge $(u, v)$, the connection between $u$ and $v$ is reliable. There is a maximum latency during $u$ sending the message and $v$ receiving \& processing the message, defined as $L_{max}$. So, all possible delays for passing messages on edges must be less than $L_{max}$. Note that the computation on the CPU of local clients is much faster than the message passing by an order of magnitude, and thus we omit the CPU time in $L_{max}$. 
    \label{def:latency}
\end{definition}

\subsection{Privacy and Security}
We specifically define privacy and security problems for distributed core decomposition with the following properties:

\begin{itemize} 
\item \emph{Locality}: 
Each vertex $u$ in the graph $G$ is only aware of its directly connected neighbours $u.\adj$, and $u$ cannot know any other indirectly connected edges. That is, the connected edges of a vertex themselves are privacy. This can be ensured by the locality of the distributed core decomposition algorithm (see Theorem~\ref{th:locality} above). 

\item \emph{Decentralized}:
Each vertex $u$ in the graph $G$ can contain only local features that are generated by itself and collected from its neighbours. No vertices can collect information from non-neighbours. 
In other words, there is no centralized server that can store the global information of $G$.
This is ensured by the decentralization of the algorithm (see Definition~\ref{def:decentral} below).



\item \emph{Secure Core Number Comparison}:
Each vertex $u$ in the graph $G$ can only exchange information with its neighbour $v\in u.\adj$ to compare the core numbers. Practically, $u$ only needs to know the Boolean result of comparing two core numbers, rather than the value of $v.\core$.  
We can minimize the leakage of private information during the comparison (see Definition~\ref{def:secure-corenumber} below).


\item \emph{Secure Calculation of Core Number Distributions}: 
After all core numbers $u.\core$ are computed for all $u\in V$, there is an issue of how to release the core numbers.
Practically, users only want to obtain the statistic result of core numbers, rather than all the exact values of $u.\core$ for all $u\in V$. 
For example, users want to know how many vertices $v$ with the same core number $1$ and label $A$, denoted $\{v\in V: v.\core = 1 \land v.\lb = A\}$; according to this result, users cannot infer the exact core numbers of any vertices (see Definition~\ref{def:result} below).

\end{itemize}

\begin{definition}[Decentralization]
    Given a graph $G$, there does not exist a central server to store all the information from the clients globally. Each vertex $v\in V(G)$ only stores the local and neighbours' information. Synchronization is implemented by each vertex communicating with its neighbours. Therefore, any vertex or server cannot collect the information of all the vertices. 
    \label{def:decentral}
\end{definition}

To identify the termination of core decomposition, we must detect whether all clients finish the computation at the same time or not. Since there is no global server to monitor all clients, we design a mechanism to do this with a global server. 

\begin{definition}[Secure Core Number Comparison]
    Given a vertex $u$ and its neighbour $v\in u.\adj$ in the graph $G$, $u$ must know its core number $u.\core$ and want to compare it with $v.\core$. The expected result for $u$ is the Boolean value of $u.\core \geq v.\core$, and $u$ cannot know the exact value of $v.\core$.
    \label{def:secure-corenumber}
\end{definition}
Suppose $k = v.\core$ is in the range $[0, k_{max}]$, and $u$ wants to know $k$. So, $u$ must compare at most $\lceil \log{k_{max}}\rceil$ times to obtain the value of $k$. 
Actually, $k_{max}$ is equal to $v.\Deg$ before the core decomposition, but $u$ cannot know $v.\Deg$. 
For example, when we know $k=v.\core$ is in $[0,10]$, we have $3$ binary search comparisons to know $k = 1$. 

\begin{definition}[Secure Calculation of Core Number Distributions]
   In a graph $G$, we assume that all the core numbers of the vertices $v.\core$ for all $v\in V(G)$ have been calculated. 
   Given a core number $k$ and a label $\lb$, denoted $(k, \lb)$, we want to count the number of vertices in the set $\{v\in V(G): v.\core = k~\land~ v.\lb = \lb\}$, denoted $n_{(k,\lb)}$. The expected result is the size of this set $n_{(k,\lb)}$, and the exact values of $v.\core$ and $v.\lb$ for all $v\in V(G)$ are not released. Also, only the single root vertex $r$ that initiates the calculation can get the result, and all other vertices cannot know the result. 
    \label{def:result}
\end{definition}

For example, a root vertex $r$ starts the calculation for the number of vertices specifically with the core number $2$ and the label $A$, $|\{v\in V(G): v.\core = 2~\land~ v.\lb = A\}|$, and the result is $1,000$; only the root vertex $r$ that initiates the counting process can know the result, and all other vertices cannot know the result. According to $n = 1,000,000$, we know that there exist only $0.1\%$ of such vertices that have $(A,2)$. However, the exact values of the core numbers for all the vertices in $V$ are kept as privacy. 



\section{Our Technical Implementation} \label{Ours}
In this section, we describe the details of the parts in the framework shown in \myfig~\ref{fig:framework}. 

\subsection{Comparing Core Numbers by HE}
\label{OursComparing}

We use asymmetric HE to compare the core numbers of two vertices. 

\paragraph{Algorithm} 
Given an edge $(u,v)\in E(G)$, we have a \emph{source} vertex $u$ and a \emph{target} vertex $v$. The vertex $u$ wants to compare its core number with $v$. In other words, $u$ wants to acquire the result of $u.\core > v.\core$.
The whole process is to extend the core numbers comparing in secure (line 13 in Algorithm~\ref{alg:core-decompotion}), and the array $u.A$ will store the result of core number comparison instead of the value of neighbours' core numbers (lines 3 and 6 in Algorithm~\ref{alg:core-decompotion}).  
The secure comparison algorithm has four steps: 
\begin{enumerate}
    \item The target vertex $v$ send the message $m_1$ to notify $u$ that $v.\core$ is decreased (or new assigned for initialization, line 4 in Algorithm~\ref{alg:core-decompotion}), where $m_1$ includes the ID of $v$ denoted as $m_1=\langle v \rangle$. 
    
    \item The source vertex $u$ generates a pair of keys $(\pri, \pub)$; we define the public encrypt function $E_{\pub}()$ and the private decrypt function $D_{\pri}()$.
    Then, the source vertex $u$ sends the message $m_2$ to $v$, where $m_2$ includes the public key and its encrypted core number denoted as $m_2 = \langle \pub, E_{\pub}(u.\core)\rangle$.

    \item The target vertex $v$ receives the message $m_2$ from $u$. Then, $v$ generates $E_{pub}(v.\core)$. According to the definition of HE, we have $E_{pub}(u.\core) > E_{pub}(v.\core) = E_{pub}(u.\core > v.\core)$. Thus, $v$ can reply the message $m_3$ to $u$, where $m_3$ includes the encrypted comparison result denoted as $m_3 = \langle E_{pub}(u.\core > v.\core)\rangle$.

     \item The vertex $u$ receives $m_3$ from $v$. Then $u$ can decrypt the message $D_{\pri}(m_3)$ with its private key $\pri$ and get the result of $u.\core > v.\core$. 
\end{enumerate}

\iftrue
\begin{figure}[tb]
    \centering
    \includegraphics[width=\figscale\linewidth]{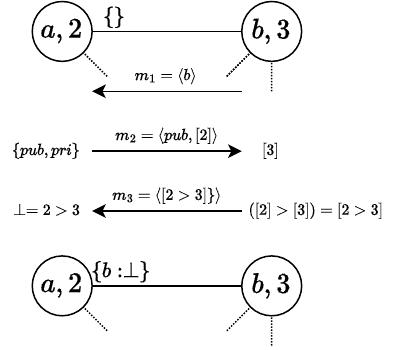}
    \caption{An example of comparing core numbers between the vertices $a$ and $b$ in \myfig~\ref{fig:distributed}(a). The arrows are the direction for passing messages between two vertices. The $[2]$ is the encrypted value $2$ with the public key $\pub$. 
    }
    \label{fig:compare}
\end{figure}

\begin{example}
In \myfig~\ref{fig:compare}, we show an example of how to compare the core numbers for the source vertex $a$ and the target vertex $b$.
The vertices $a$ and $b$ have the core numbers $2$ and $3$ initialized by their degree, respectively.
First, $b$ sends $m_1$ to $a$, indicating that $b.\core$ is updated. 
Second, $a$ generates a pair of public and private keys; $a$ sends the public key $pub$ and the encrypted core number $2$ to $b$. 
Third, $b$ sends the encrypted result for the core number comparing $2 > 3$, to $a$. 
Finally, $a$ decrypts the result and gets the result to be \texttt{false}, which is stored in the array $A$ for later updating $a.\core$.
\end{example}
\fi 


\paragraph{Privacy Security Analysis}
In the end, only the source vertex $u$ knows the result of $u.\core > v.\core$, but $v$ does not know this result; also, $u$ cannot know the value of $v.\core$, and vice versa. However, there are two cases where the values of core numbers are leaked. 

First, given $v.\core$ is in the range $[0, k_{max}]$, $u$ can compare $\lceil\log k_{max}\rceil$ times to acquire the value of $v.\core$, by using a binary search. That is, at first, $u$ can pretend that its core number is $k_{max}/2$, compare with $v.\core$, and identify the half range in which $v.\core$ is. This process continues until $u$ acquires the value of $v.\core$. 
To solve this problem, a straightforward solution is that the target vertex $v$ can limit the number of comparisons for the source vertex $u$. In this case, we can ensure that $u$ only compares the core number with $v$ once, with minimal leaking of the value of $v.\core$ to $u$. 

Second, a special case is that if we have $u.\core = 2~\land~u.\core > v.core$ as true, we can infer that $v.\core = 1$, since $v$ must be connected to $u$ and $v.\core$ cannot be $0$. In this case, the source vertex $u$ can infer that $v.\core=1$, which is the minimal leaking of the value of $v.\core$ to $u$.

\paragraph{Message Complexities} 
The whole process involves three messages, $m_1$, $m_2$, and $m_3$, so that the source vertex $u$ can compare the core number with the target vertex $v$. 
Obviously, the length of $m_1$ is bounded by $O(\log n)$ since it only includes the ID of the vertex. 
The length of $m_2$ is bounded by $O(\log n)$, since the encrypted Boolean value always has a fixed length. 
The public and private keys have fixed length, e.g., the RSA typically ranges from 1024 bits to 4096 bits. So, $m_2$ includes the public key $pub$, where the length is bound by $O(\log n)$. Therefore, $m_1$, $m_2$, and $m_3$ need not be split, and three messages are required for our one-time secure comparison of the core numbers. 

Compared with our method, in Algorithm~\ref{alg:core-decompotion}, $v$ only needs to send one single message that includes $v.\core$ to $u$ when $v.\core$ is updated, where the core number must be bounded by $O(\log n)$. 
In other words, our secure algorithm spends triple messages to ensure the protection of the core numbers.

\paragraph{Efficiency}
It is true that HE-based comparisons and key exchanges are slow and have high computational cost. However, the HE-based computation must be performed on local clients; in distributed networks, the message passing is much slower than HE-based computation since it has latencies (like 10ms). Therefore, practically, the main performance bottleneck is message passing rather than HE-based operations.

\subsection{Decentralized Termination Detection}
\label{OursTermination}
We describe how to detect the termination of the distributed core decomposition algorithm without using a centralized server, that is, a decentralized approach. Our approach is based on the Feedback-BFS tree and heartbeats.
We first define that each vertex $u$ needs to maintain a status, denoted $u.s$, which has two values: 
\begin{itemize} 
    
    \item \live: the vertex $u$ actively receive the messages from neighbours $u.\adj$ and process these messages; or $u$ actively send the messages to neighbours $u.\adj$. Initially, the vertex $u$ must be \live\ since it begins to send $u.\core$ to all neighbours in $u.\adj$. 

    \item \dead: the vertex $u$ stops calculating the core number since the Locality of $u$ (Theorem~\ref{th:locality}) is satisfied; and $u$ is not receiving or sending messages.

\end{itemize}
In Algorithm~\ref{alg:core-decompotion}, before lines 2 and 6, we insert $u.s \gets \live$, respectively; after lines 4 and 11, we insert $u.s \gets \dead$, respectively. This is to explicitly show the status of the current vertex.  

\begin{theorem}[Termination Condition]
    The distributed $k$-core decomposition is terminated if the status $u.s$ for all vertices $u\in V$ is \dead\ simultaneously within a period that must be greater than $L_{max}$.
    \label{th:termination}
\end{theorem}
\begin{proof}
From Algorithm~\ref{alg:core-decompotion}, we can see that the termination condition is that all vertices stop calculation, and also stop sending and receiving messages.
It is possible that the vertex $u$ has sent the message to $v$, but $v$ has not yet received the message due to latency up to $L_{max}$. Since such a latency can be long, we must consider it when detecting the termination. In other words, the distributed algorithm is not terminated if there exists at least one \live\ vertex for a period greater than $L_{max}$. 

\end{proof}




\paragraph{Feedback-BFS Tree}
\label{sec:bfs-tree}
Given a graph $G$, we first select a root vertex $r$ to build a Feedback-BFS tree. The process is straightforward with four steps: 
\begin{enumerate}
    \item Starting from the root vertex $r$, it sends the message $m_1$ to all neighbours $r.\adj$, where $m_1$ includes a version number $\ver$ in case there are multiple BFS trees generated on the same graph, denoted $m_1 = \langle \ver \rangle$. Here, the root vertex has {children} without {parent}. 
    
    \item The vertex $v$ receives the message $m_1$ from a neighbor $u'$. If $v$ is the first time receiving $m_1$ (the parent $v.\pr = u'$), it forwards $m_1$ to other neighbours $v.\adj \setminus u'$ who have not received $m_1$ (the children $v.\ch$). 
    This process will be repeated until $v$ is a leaf vertex that does not have children.
    
    \item The leaf vertices $w$ ($w.\ch=\emptyset$) receive $m_1$ and will reply the message $m_2$ to its parent $w'$, where $m_2$ includes an opposite version number $-\ver$ in $m_1$ (indicate that $m_2$ is a corresponding replied message), denoted $m_2 =\langle -\ver \rangle$. Then, $w'$ will reply to its parent $w'.\pr$ at once if all children $w'.\ch$ have received $m_2$. This process will be repeated until the root vertex $r$ receives the message $m_2$. 
    

    \item Finally, we obtain the Feedback-BFS duration $\overline T$, which is the time period between the root vertex $r$ sending $m_1$ and receiving $m_2$ (see Definition~\ref{def:bfs-duration}). 
\end{enumerate}
\begin{definition}[Feedback-BFS Duration $\overline T$]
Given a root vertex $r$ with a BFS tree generated in the graph $G$, $r$ sends messages to its children to generate the BFS tree at time $t_1$. Then $r$ receives all the feedback messages from the children at time $t_2$. Here, a vertex $v$ in the graph can send a feedback message to the parent only if $v$ receives all feedback messages from the children recursively. The BFS duration is denoted as $\overline T = t_2 - t_1$ in a round way. That is, any vertex can receive the message from the root $r$ within $\overline T/2$  in one way.
\label{def:bfs-duration}
\end{definition}

 We can see that this BFS tree generates the fastest path from the root to all the other vertices. Since message passing is slow, the first message that a vertex $v$ can receive must come from the fastest path. In other words, the duration $\overline T$ means the round time between sending messages to the furthest vertex and receiving the feedback. Our termination detection is based on this BFS tree.

\begin{theorem}[Reachable in $\overline T$]
Given a generated Feedback BSF tree on a distributed graph $G$, we can select any pair of vertices $(u, v)$. The vertex $u$ can broadcast the message $m_1$ starting from $u$, and $v$ can always receive $m_1$ within the duration $\overline T$.  
\begin{proof}
Suppose that the root vertex is $r$. The vertex $u$ can broadcast the message $m_1$ to $r$ within $\overline T/2$; and the $r$ can continuously forward the message $m_1$ to $v$ within $\overline T/2$. Therefore, the total time for $u$ can broadcast $m_1$ to $v$ costs at most $\overline T$ time. 
\end{proof}
\end{theorem}

\begin{figure}[tb]
    \centering
    \includegraphics[width=\figscale\linewidth]{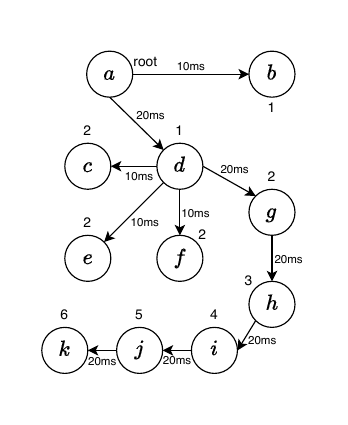}
    \caption{An example of generating a BFS tree. The root vertex is $a$. The numbers beside the circles are the depth of the vertices from the root. The arrows of edges indicate the direction of broadcasting the messages, which can only be from the parent to the child. The time beside the edges is the latency of the message passing via the edges. }
    \label{fig:bfs-tree}
\end{figure}

 \begin{example}
    An example is shown in \myfig~\ref{fig:bfs-tree}. A BFS tree is built by choosing the vertex $a$ as the root. 
    The vertex $c$, $e$, $f$, and $k$ are leaves without children; the vertex $d$ has a parent $a$ and children $c,e,f$ and $g$; the root $a$ does not have a parent but has two children $b$ and $d$.The BFS duration $\overline T$ is the round time of the message passing between $a$ and the furthest leaf vertex $k$, where we have $20 \times 6 = 120$ ms for one way and 240 ms for the round way. We observe that there are totally $11$ edges given $12$ vertices in this BFS tree.
\end{example} 

\paragraph{Heartbeat-Based Approach}
After building the BFS tree for a given graph, each vertex sends the heartbeat message to notify other vertices that it is \live. Each vertex will send the heartbeat to other vertices in the BFS tree. We first define two attributes, \emph{interval} and \emph{duration}, for the heartbeat:

\begin{itemize}
    \item {Heartbeat Interval $I$}: Since our system can accept the latency of detecting termination, the heartbeat interval can be large enough to ensure low network resource consumption, but it also should be much less than the total running time. 
    
    \item {Timeout Duration $T$}: Typically, $T$ is set to a multiple of the interval $I$ (e.g., 2 - 3 times the interval). This gives some leeway for occasional delays in heartbeat delivery. For example, we can set $I$ as 10 s, giving T as $30$ s. 
\end{itemize}

\begin{theorem}
Given the Feedback-BFS duration $\overline T$, we can choose the Timeout Duration $T = 3\overline T/2$, and the Heartbeat interval $I = T/3$. 
\begin{proof}
We choose the value of $T$ for two reasons. First, we must consider the time that the message passes through networks, that is, a heartbeat can be received by other vertices using at most $\overline T$ time. Second, we must have some leeway for the timeout duration for measuring the time, e.g. $1/2$ of $\overline T$. 
\end{proof}
\end{theorem}

Based on the determined $T$ and $I$, we define a second status $u.s'$ for each vertex $u$ to detect termination, which includes two values: 
\begin{itemize}
    \item \Active: the vertex $u$ always receives at least one heartbeat within the Timeout Duration $T$; or $u$ is computing, that is $u.s=\live$. 

    \item \Inactive: the vertex $u$ does not receive the heartbeat within the Timeout Duration $T$; and $u$ is not computed, that is, $u.s=\dead$.
\end{itemize}

Only the \live\ vertices $u$ that perform computations with $u.s = \live$ can generate and send heartbeats. 
The detailed process is as follows: 

\begin{enumerate}
    \item Initially, all the vertices $u\in V$ are set to \live\, which have the $u.s'$ set to \Active.  
    
    \item For each vertex $u\in V$, if $u$ is executing the computation with $u.s=\live$, $u.s'$ is set to \Active; then, $u$ will continuously send heartbeats to neighbours in the BFS tree;
    otherwise, $u.s'$ is \dead\ and stop sending heartbeats.

    \item For the vertex $v$, if $v$ receives the heartbeats from $u$ within $T$, $v.s'$ is set to \Active; then, $v$ will continuously forward heartbeats to other neighbours in the BFS tree except $u$;
    otherwise, $v.s'$ is set to \Inactive\ and stop forwarding heartbeats.  

    \item Steps 2 and 3 will repeat until the heartbeats are broadcast to all vertices in the graph using the BFS tree.
    
\end{enumerate}

By doing this, we only need to check the status of $u.s'$ for any vertices $u\in V$. If $u.s'$ is \Inactive, we know that the distributed core decomposition is terminated; otherwise, the algorithm is still running. In other words, any vertices $u$ always have the status $u.s'$ indicating termination.

\begin{example}
As an example BFS tree shown in \myfig~\ref{fig:bfs-tree}, we suppose that only the vertex $h$ is \live\ and all other vertices are \dead\ as shown in \myfig~\ref{fig:distributed}(d).

In the beginning stage, all vertices are set to \Active. 
Only the vertex $h$ is doing the computation, so $h$ is \live; it immediately generates heartbeats and sends them to all its BFS neighbours, $g$ and $i$. These heartbeats will be repeatedly forwarded to all other vertices, which are set to \Active. For example, $g$ is set to \Active\ when receiving heartbeats from $h$ after 20 ms, then $g$ immediately sends heartbeats to $d$ and $d$ is set to \Active\ when receiving them after 40 ms. 
In this case, the fast vertex $b$ will receive the heartbeats and be set to \Active after 70 ms.
Finally, all vertices can receive the heartbeat and be set to \Active. 
That is, the \live\ vertices can only generate and send heartbeats; the \dead\ vertices can only forward the heartbeats generated by the \live\ vertices.

We choose the Timeout Duration $T = 360$ms, larger than the BFS Duration $\overline T = 240$ms. So, we choose the Heartbeat Interval $I = 360/3 = 120$ ms. 
We can test the termination with any vertices, e.g. $b$; that is, if $b$ cannot receive heartbeats within $T$, $b$ is set to \Inactive. We detect that the algorithm has been terminated.

\end{example}

\paragraph{Privacy Security Analysis.}
For the entire process, the root vertex is selected arbitrarily, and any vertex can be the root, which is not centralized. Each vertex $u$ must record the information of the neighbours $u.\adj$ for the parent and children in the BFS tree; and $u$ does not know any other connections of the neighbours. Finally, the root vertex can know the BFS duration $\overline T$, which is the information of the whole graph; but the latency for a specific edge cannot be inferred from $\overline T$. 

\paragraph{Message Complexities.}
We only analyze the Feedback-BFS tree.
Initially, the vertices send a total of $m$ messages to build the BFS tree. The BFS tree has $n-1$ edges, so the vertices respond with $n-1$ feedback messages. Therefore, in the worst case, the total number of messages is $\mathcal W = O(m+n)$, and the depth is the longest chain in the graph, denoted as $\mathcal D = O(|D|)$. 


\subsection{Calculate the Distribution of Core Numbers}
\label{OursRelease}
After the termination of the distributed $k$-core decomposition, we need to release the core numbers as the results of the computation to minimize privacy leakage.  

\paragraph{Algorithm}
Given a graph $G$, we assume that the BFS tree is already built with the root vertex $r$, which is used for the detection of decentralized termination. Then, for the vertex $u$, we select a label and a core number, denoted $u.(\lb, k)$, to count the number of such vertices, denoted $n_{(\lb, k)}$. 
The detailed process is as follows:

\begin{enumerate}
    \item Starting from the root $r$, it generates a pair of keys $(\pri, \pub)$ with the public encrypt function $E_{pub}$ and the private decrypt function $D_{pri}$. It sends the message $m_1$ to all children $u \in r.\ch$, where $m_1$ includes the public key, a pair of the label and the core number, and a version number $\ver$ in case of multiple instances of the BFS tree, denoted $m_1 = \langle \pub, E_{\pub}(\lb, k), \ver\rangle$. 

    \item The vertex $u$ receives the message $m_1$. If $u$ has children $u.\ch$, $u$ will forward $m_1$ to all $u.\ch$. This process will repeat until all vertices $u\in V$ receive $m_1$. 
    
    \item The leaf vertices $u$ ($u.\ch = \emptyset$) receive $m_1$ and will reply to the message $m_2$ to its parent $u.\pr$. If the vertex $u$ has the same label and core number as in $m_1$, denoted $E_{\pub}(u.(\lb,k)) = E_{\pub}(\lb, k) \in m_1$, we reply $m_2$ to $u.\pr$, where $m_2$ includes the encrypted counting number $1$ and the opposite version number $-\ver$ (indicate that $m_2$ is a corresponding replied message), denoted $m_2 = \langle E_{pub}(1), -\ver\rangle$; otherwise, the replied message $m_2=\langle E_{pub}(0), -\ver\rangle$. In other words, the leaf vertices start to count.


    \item The non-leaf vertices $u$ ($u.\ch \neq \emptyset$) receive the message $m_2$. If all children $v\in u.\ch$ receive $m_2$ together, $u$ will add all $E_{\pub}(n_{(\lb,k)})$ of the children together and reply to $u.\pr$ with the message $m_2 = \langle E_{\pub}(n_{(lb, k)}), -\ver\rangle$. Here, HE is used for the encrypted add operation like $E_{\pub}(1) + E_{\pub}(2) = E_{\pub}(3)$. 
    
    \item This step (3) and (4) will repeat until all children of the root vertex $u\in r.\ch$ receive the message $m_2$. Finally, the root $r$ will obtain the complete number of vertices with the corresponding $(\lb, k)$. With its local private key, the root $r$ can decrypt the counting, denoted $n_{(\lb, k)} = D_{\pri}(E_{\pub}(n_{(lb, k)}))$. Now, we obtain the result of $n_{(\lb, k)}$. 
    
\end{enumerate}



\begin{figure}[tb]
    \centering
    \includegraphics[width=\figscale\linewidth]{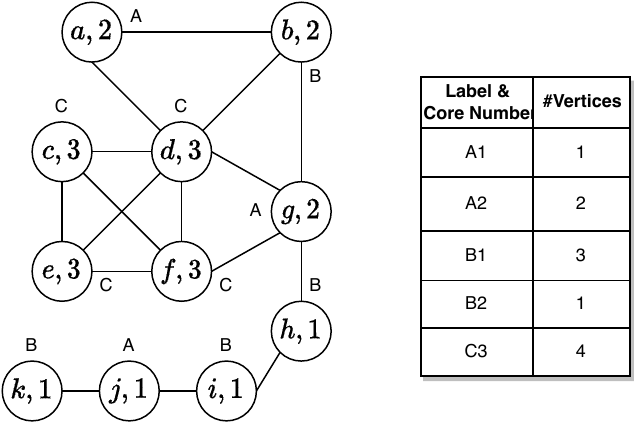}
    \caption{
    An example of calculating core number distribution. 
    The capital letters beside the circles are the labels of the vertices. The right table shows the result of core number distribution, where the first column shows the labels and core numbers, and the second column shows the number of vertices.
    }    
    \label{fig:result}
\end{figure}

\begin{example}
In \myfig~\ref{fig:result}, we show an example of calculating the distribution of the core numbers. After core decomposition, we observe that the vertices have core numbers and different labels, $A, B$ and $C$. 
We want to release the number of vertices with the label and $B$ and the core number $1$, which is shown in the right table $B1$ with $3$.

Using the BFS tree shown in \myfig~\ref{fig:bfs-tree}, the root $a$ sends the message $m_1$ to notify all vertices. The leaf vertices, $b,c,f$ and $k$, will first reply the message $m_2$ to their parent vertices, $a,d$ and $j$. This process repeats until the root $a$ receives all the messages $m_2$ from its children, $b$ and $d$, where $b$ has counter $0$ and $d$ has counter $3$. Therefore, we obtain the final counter $3$ for $B1$ as the result.

During this process, we can see that $n_{(B,1)}$ in $m_2$ is encrypted by HE. Thus, only the root vertex $a$ can acquire the corresponding values, and all other vertices can only perform the calculation without knowing the values. 
\end{example}


\paragraph{Privacy Security Analysis}
For the entire process, the root vertex is selected arbitrarily, and any vertex can be the root, which is not centralized. In addition to the root $r$, each vertex $u$ performs the add operation on the number of vertices with ${E_{\pub}(n_{(\lb, k)})}$ for his children, but $u$ does not know the specific value due to encryption. 
In addition, $u$ does not know if it is counted or not in $n_{(\lb, k)}$, since $u.(\lb,k)$ is encrypted for comparison.
Therefore, $u$ only knows whether the calculation of the core number distribution is executed or not. But $u$ does not know which root vertex $r$ launches this calculation and which pair of labels and core numbers $(\lb, k)$ is counted. Only the root $r$ has the private key for decryption and can get the final result of $n_{(\lb, k)}$, so that all the other vertices cannot know such a result.

\paragraph{Message Complexities}
When releasing one result for one pair given a label and a core number, since the BFS tree has $n-1$ edges, the total number of messages is $\mathcal W = O(2(n-1)) = O(n)$. The depth is the longest chain in the graph, denoted as $\mathcal D = O(D)$.

For releasing the results for all pairs given labels and core number, in the worst case, we can have a maximum number for the combination of labels and core numbers up to $|\mathcal L|\cdot k_{max}$. So, the algorithms must run at most $|\mathcal L|\cdot k_{max}$ to obtain all of the core number distributions.
Typically, we can choose different root vertices to start the calculation of the core numbers at the same time. It can be executed in parallel with high probability, as most of the running time is spent on the message passing, and the calculations on clients are much faster than the latency of messages.

\section{Conclusion and Future Work}
\label{Conclusion}
In this work, we investigate the classical distributed $k$-core decomposition algorithm. Then, we adequately solve the privacy and security problems in terms of comparing core numbers, decentralized termination, and releasing core number distributions. 

In the future, first, we will explore the possible adversarial behaviours. e.g. the false heartbeats in terminal detection. 
Second, we will conduct experimental evaluations of our methods on various real and synthetic data graphs, e.g., billions of vertices \& edges social networks, hyperlink networks, or biological networks. However, it is hard to build a distributed environment that may require billions of clients. We can simulate a great number of clients on a signal machine, e.g. thousands of goroutines can run simultaneously synchronized by message passing by using Golang~\cite{guo2024experimental}. 
Third, we will explore secure distributed $k$-core maintenance algorithms~\cite{guo2022simplified,guo2023parallel}in dynamic graphs, in which edges are continuously inserted or removed, and the core numbers can be efficiently maintained instead of recalculating.
Finally, we can apply our methodology to other distributed graph algorithms, e.g. distributed $k$-core maintenance~, to protect privacy and security. 

\bibliographystyle{SageH}


\bibliography{ref.bib}
\end{document}